# Challenges of Applying E-Learning in the Libyan Higher Education System

Entisar Alhadi Al Ghawail[1], Sadok Ben Yahia[2], Mohamed A. Alrshah[3]
[1]Faculty of Sciences, University of Tunis El Manar, Tunisia, ent1977@yahoo.com
[1]Faculty of Information Technology, Al Asmarya Islamic University, Zliten, Libya.
[2]Tallinn University of Technology, Estonia, sadok.ben@taltech.ee
[3]Universiti Putra Malaysia, Selangor, Malaysia, mohamed.asnd@gmail.com

**ABSTRACT**

The adoption of ICT in classrooms is very important in order to improve education quality, promote effective management of knowledge, and improve delivery of knowledge in higher education. Some of the Libyan universities have already started using E-learning in classrooms, but many challenges are still hindering that adoption. This paper endeavors to find the obstacles that may face the adoption of E-learning in Libya and sketches out the possible solutions. Further, it highlights the potentials for the adoption of E-learning in the higher education system in Libya using both qualitative and quantitative approaches. Both questioner and interview have been used on a focused group to collect the data. Teachers and students at Al Asmarya Islamic University have been selected as a sample for this study. This paper reveals that the challenges hindering teachers and students from using ICT and E-learning are: the lack of knowledge about ICT and E-learning, the lack of ICT infrastructure, and the lack of financial support. However, the participants show a high level of interest in applying the ICT and E-learning in the university despite the unsuitability of the environment.

**1. INTRODUCTION**

Recently, Information and Communication Technologies (ICTs) including computers, accessibility to Internet, broadcasting media such as radios or TVs, and telephony have been used for educational purposes [1]. In [2] indicates that there is a need to adopt E-learning and ICTs in Higher Education in Libya. Moreover, the Libyan Ministry of Higher Education and Scientific Research has started to rebuild the Libyan education system and enhance the adoption of E-learning. Consequently, there is an urgent need to carry out many studies about the use of E-learning in Libyan Universities to extend knowledge and to fill the gap in the literature review. Although Libya aims to play a significant role among African countries through encouraging and enhancing ICT projects, non-mature infrastructure and less of qualified ICT teachers can be considered as the major challenge to the reform process [3]. Because of the civil war and ongoing conflict in Libya since 2011, no attention has been paid by the government to improve the use of E-learning and ICT in the Libyan Higher Education System (LHES). Therefore, research about the implementation of E-learning in the higher education context in developed countries is vital. This study aims to discuss the reasons behind the weak use of ICT in the LHES.

The objectives of this paper are:
1. To assess the adoption of E-learning and ICT in the Al Asmarya Islamic University (AIU).
2. To highlight the challenges that may hinder the success of adopting E-learning and ICT.
3. To outline the potential solutions that can be followed in order to address these challenges.

This study uses both qualitative and quantitative approaches since this is the most appropriate approach for this kind of research [4]. This is in order to explore the participants' perceptions and perspectives. Teachers and students from the AIU have been targeted by the questionnaire and interviews in order to collect the data. The key that we would like to address are the following:
1. What is the significance of using E-learning and ICT in the LHES?
2. To what extent the use of the Internet by teachers and students can affect the education quality in Libya?
3. What are the major challenges that may hinder the success of adopting E-learning and ICT in LHES?
4. What are the possible solutions that can be followed to overcome these challenges?

The remainder of this paper is organized as follows. Section 2 presents the operational definitions that will be used throughout the paper. The literature review about the challenges of E-learning and ICT in the LHES is introduced in Section 3 while the methodology is presented in Section 4. The experimental results and discussion are presented in Section 5, and the conclusion is exhibited in Section 6.

## 2. OPERATIONAL DEFINITION

The ICT is a lot of innovative devices and assets that are useful to convey, store and oversee data [4]. Also, ICT has been an umbrella term that incorporates any PC, hand held gadget, and intelligent white board, the Internet, applications (such as database and graphical applications) and broadcasting device [5].

**E-learning**: It is a learning that is conducted using an application or tool to help and encourage individuals on the learning, for example, online courses, online exams, online cooperation, remote assistance. In [6] categorizes the E-learning as the conveyance of preparing content by means of any electronic medium, while the reference [7] defines the E-learning as teaching or learning using ICT.

**LHES**: includes both public and specialized universities as well as higher professional institutes. The latter provide training of teachers while the higher institutes provide training for trainers and instructors. As for the higher educational institutes and polytechnic institutes provide industrial, technical and agricultural sciences support. Additionally, few scientific research centers have been established to support health, pharmaceutical, educational, and basic sciences [8].

**Smart Learning Environments**: It is an ICT-based learning solution, which is highly integrated into an educational institute environment. Therefore, a smart learning environment is a physical environment enriched with environment-aware ICT devices to enhance and speed up the learning process [9]–[11].

## 3. LITERATURE REVIEW

In this section, the literature review has been presented, including the current situation in LHES and their plans of implementing ICT and E-learning, and the Webometrics classification for the Libyan universities.

### 3.1 Teaching and Learning by ICT

ICT is of extensive use in tertiary institutions to facilitate teaching, by which ICT could change the teaching method from face-to-face teaching into either blended or online learning [12]–[16]. This process has the ability to make courses available and accessible online to help students take or retake the course at any time. According to [17], face-to-face can be used without E-learning, but in this case, it will be called blended learning due to the use of classroom aids such as computers and internet in the classroom. The authors of [14] and [16] argue that asynchronous learning happens where the interaction between teacher-students and students-students takes place at the same times and in the same online space. Needless to mention that in the LHES, ICT is an essential element that can help both learners and instructors.

### 3.2 The LHES and the implementation of the ICT

The LHES has the potential to enhance the economic growth through knowledge economy and human capital. Libya has started to rehabilitate its LHES by attempting to use E-learning and ICT to deliver knowledge. However, this requires to create an environment where new ways of learning are encouraged and accepted [3]. The LHES policy for applying the ICT in education firstly aims to increase accessibility to the ICT applications and tools. Therefore, they have to rebuild the infrastructure to support ICT requirements to guarantee the best possible learning tools such as devices, materials, strategies, and media to cope with the advances that exist in other countries.

However, because of the civil war and ongoing conflict in Libya, the internet access is often weak or interrupted and the academic staff needs to be trained to use ICT in order to deliver knowledge and skills online. According to [18], Libyan universities need to utilize ICT in classrooms to cope with the improvements prodded by mechanical progressions. Besides that, some key factors are yet disturbing the comprehension of ICT, which are the absence of ICT skills and the lack of hierarchical support. Table 1 summarizes the significant challenges that can hinder the success of E-learning implementation in developed countries [21].

**Table 1:** The challenges facing the application of E-learning in developed countries [2], [5], [6], [18]–[20]

| Ref No. | Country | Challenges |
|---|---|---|
| [18] [20] | Libya | - Internet low connection.<br>- Lack of aid.<br>- Lack of language proficiency.<br>- Lack of courses for training.<br>- Lack of trained academic staff.<br>- Cost of internet.<br>- Social restrictions and disinterest.<br>- The instability and insecurity because of the ongoing conflict. |
| [6] | Jordan | - Lack of institutional support and encouragement.<br>- Undeveloped infrastructure. |
| [2] | Saudi Arabia | - Internet cost.<br>- Lower esteem for public of web-based. |
| [21] | Iraq | - Lack of qualified and prepared scholarly staff<br>- Lack of specialized help.<br>- Lack of ICT framework.<br>- Lack of mindfulness, premium, and inspiration toward E-learning innovation.<br>- The postponement of acquainting E-learning innovations with advanced education.<br>- Low web data transfer capacity.<br>- Insufficient money related help.<br>- Inadequate preparing programs.<br>- Electricity defficiency, ICTs and E-learning absence of education. |
| [4] | Kenya | - Lack of budgetary help.<br>- Lack of moderate and sufficient Internet data transfer capacity.<br>- Lack of operational E-learning strategies. |

### 3.3 Classification of Libyan universities according to their web content

It has been argued that the Webometrics classification is the most known classification of international universities, which ranks those universities based on their contents on the Internet. This classification is issued every six months by the Cyber-metrics Lab of the Spanish Supreme Council for Research (CISC) in Spain. This classification aims to enhance competition among universities in order to increase the content of electronic scientific materials on the Internet. The classification is based on four basic criteria: clarity; effectiveness; attendance; and openness. Figure 1 shows the ranking of Libyan universities according to Webometrics.

Reference [18] shows that E-learning can possibly give learning openings everywhere and at any time since it is probably going to empower students and educators to use the web to convey, share assets, and open instructive chances. Additionally, ICT and E-learning can be utilized as reconstructive and appealing measures to help Libyan students and educators. Moreover, it makes the learning condition as a context to pass information from instructors to students, from students to educators, and from students to students, and as a spot for innovative reasoning and learning. As well as, it has been expressed that the use of ICT and E-learning is still in its beginning stage in Libya even before the fall of Gaddafi's regime [8].

| World Ranking | University Name | World Ranking | University Name |
|---|---|---|---|
| 4198 | University of Benghazi (University of Garyounis) | 21712 | College of Medical Technology Misurata |
| 4569 | University of Tripoli (Al Fateh University) | 22804 | College of Computer Technology Zawiya |
| 5311 | Misurata University (Misrata University) | 23235 | College of Electrical and Electronic Technology Benghazi |
| 5414 | Libyan International Medical University | 24158 | University of Tripoli Alahlia for Humanities and Applied Sciences |
| 6191 | Sebha University /Sabha University | 25320 | Libyan Institute for Advanced Studies |
| 6364 | Omar Al Mukhtar University | 25370 | College of Industrial Technology Masrath |
| 9786 | Sirte University (Al Tahadi University) | 25978 | College of Computer Techniques Tripoli |
| 11393 | Al Zawiya University (Seventh of April University) | 26089 | Azzaytuna University |
| 12252 | University of Elmergib | 26615 | Sabratha University |
| 13575 | Al Asmarya University of Islamic Sciences | 27664 | Africa University for Human and Applied Science Tripoli City |
| 20104 | Academy of Graduate Studies | 27713 | University of Tobruk |
| 20728 | Petroleum Training and Qualifying Institute | 27797 | College of Electronic Technology Tripoli |
| 21570 | University of Gharyan | 28139 | University of Ajdabiya |

**Figure 1:** Ranking Web of Libyan Universities by Webometrics [2]

Recently, many ICT projects in Libya have been in progress and one of these projects has been cosponsored by the Libyan government and the UNESCO. This project focuses on the reform of the LHES through rebuilding local area networks within faculties in different universities' campuses and institutes. It also aims at creating a national ICT resource center such as unified student information systems.

## 4. METHODOLOGY

This study uses both qualitative and quantitative since the mixed method is the most appropriate approach for this kind of research [5], in which it explores the participants' perceptions and perspectives. Teachers and students from the AIU have been targeted by the questionnaire and interviews in order to collect the data.

The questionnaire consists of two parts: (1) general questions to collect information about the participants themselves and (2) open-ended questions to give them the opportunity to freely express their opinions about the problem of this research. Thirty (30) teachers and students from different disciplines in AIU have been targeted by both questionnaire and interviews, where ethical issues such as confidentiality and consent have been taken into consideration. Thereafter, SPSS has been used to analyze and interpret the collected data.

## 5. RESULTS AND DISCUSSIONS

The results shown in Table 2 reveal that the infrastructure is an essential and crucial element for the implementation of the E-learning in the LHES as the participants indicate that the infrastructure like computers, networks, Internet connectivity, and computer labs are inadequate in most Libyan universities. The results show high student interest in using ICT and E-learning, which has to be supported by the LHES.

### 5.1 Q1. The teachers' use of ICT in the LHES.

Despite that about 43% of the participants have high-level of ICT skills, 83% of the participants rely mostly on their own PCs and Internet source with limited speeds in order to improve their approaches of teaching. The results additionally show that around 60% of the participants express that they are persuaded to utilizing the ICT and E-learning in their classrooms.

**Table 2:** The availability of the main factors hindering the implementation of E-learning in AIU.

| The Phrase | Yes Frequency | Yes Percentage | No Frequency | No Percentage |
|---|---|---|---|---|
| Infrastructure is available. Laboratories are available. There are enough computers. | 8 | 27% | 22 | 73% |
| Internet is available for teachers. | 25 | 83% | 5 | 16% |
| Teachers have enough skills to use computer and Internet. | 13 | 43% | 17 | 56% |
| Social networking applications are used to communicate with students. | 18 | 60% | 12 | 40% |
| Email is used to communicate with students. | 15 | 50% | 15 | 50% |
| There are dedicated halls for technology and E-learning. | 0 | 0% | 30 | 100% |
| Teachers are convinced of using the Internet and E-learning. | 18 | 60% | 12 | 40% |

The results likewise uncover that LHES faces numerous difficulties which have been hindering the implementation of ICT and E-learning in Libya since decades. As indicated by [22]–[26], in spite the fact that innovative ICT and E-learning tools are accessible and available, the decision makers in the LHES think that it is difficult to apply them due to their high costs and due to the absence of the administrative will, capacity and aptitudes.

### 5.2 Q2. Hampering Challenges

The results show main significant challenges which prevent the successful implementation of ICT and E-learning at AIU are as shown in Figure 2. In general, awareness of ICT and E-learning can be considered as low among academic staff in LHES since the majority of students and teachers have low or no previous knowledge about ICT, especially in the human studies, arts and Islamic fields. Moreover, the negative attitudes towards ICT affect a good number of decision-makers who are in charge of the implementation of ICT and E-learning, especially the elderly people.

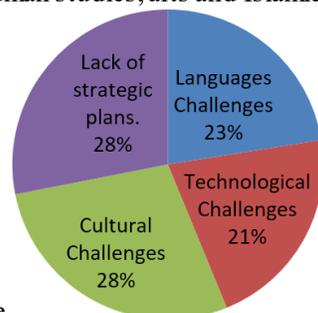

**Figure 2:** The challenges of using of E-Learning and ICT

**Language challenges**: The language barrier hinders the adoption of E-learning in Libya especially because the official language in Libya is Arabic, most of the teachers and students have no English proficiency, and most of the online materials and references have been written in the English language. About 23% of the participants confirm that the language barrier is very crucial and vital.

**Technological challenges:** ICT is the key element in the development of E-learning, including all requirements such as networks, hardware, software, computers, radio, TV, audio cassettes, video, and Internet access. Libya faces a number of challenges on the technological front because the Libyan universities appear to lack the technological infrastructure. The use of educational software within universities is very limited too because of the poor Arabic-software available in the market. Finally, technical support appears to be very poor and about 21% of the participants feel that the technical support is indispensable for the adoption of the E-learning.

**Cultural Challenges:** Any university striving to obtain a successful E-learning strategy must be prepared technologically as well as culturally. The socio-cultural factors may pose several barriers during the implementation of E-learning. Those factors such as the style of interaction and communication could have a high impact on how people learn. The results show that 28% of the participants consider those factors as an important motivational factor.

**Lack of clearly defined strategic plans and E-learning policies:** The results reveal that the lack of operational E-learning policies is another challenge that is quite likely to hinder the E-learning in AIU. It has been noticed that many Libyan public universities have no E-learning policy and even if it is available, it is not operational and this may possibly be due to the lack of ICT infrastructure. About 28% of the participants confirm that the lack or weakness of these policies is one of the most significant factors hindering the implementation of the E-learning in LHES.

### 5.3 Q3. Recommendations of possible solutions

The key elements of the future solution to implement and apply E-learning in LHES are as follows:

**Curriculum development:** It is worth to mention that the curriculum becomes more interactive and dynamic if ICT-based learning activities, projects, and software applications were involved. However, this requires the development of the curriculums to cope with those tools.

**Management Support:** It could be argued that teachers could become more motivated when they receive support and encouragement from their top managements. Institutional leaders are a determinant factor, given their decision-making roles which could either make-or-break the E-learning projects by either facilitating or impeding its implementation within their institutions. Most of the participants agreed that this factor is very crucial to adopt E-learning.

**Technology Transfer:** It appears that the development of technology assists to transfer knowledge, and this can be significant for developed countries such as Libya. The study shows that the challenges are the lack of the infrastructure and the dominance of the traditional education system in Libya, but this has moved to new means of teaching and learning. Technology has offered incredible and energizing chances to create keen learning conditions, just as the potential advantages of using those new advancements by the instructors and students to facilitate their teaching and learning process.

## 6. CONCLUSION AND RECOMMENDATIONS

This research has been done to help LHES to establish an improved E-learning environment by highlighting the main challenges that hinder the success of E-learning implementation. Results of this research show that the main challenges hindering the E-learning adoption in LHES are the language barrier, technological and cultural issues, and the lack of clear strategic plans and policies to implement E-learning.

Consequently, the recommendations of potential solutions for LHES to achieve a successful E-learning implementation are as following:

1. Finding plans to execute E-learning since it a non-existing experience in the Libyan universities. Thus, we recommend a joint effort with successful universities in other countries such as Malaysia, Singapore, China, and Japan that have achieved pushed adventures in E-learning. We also recommend paying less attention to the cooperation with the neighboring countries which also suffers from the same issues.

2. Creating security plans in all Libya cities to provide protection for the ICT infrastructure, which is indispensable to guarantee the successful implementation of E-learning.
3. English proficiency programs provisioning to improve language skills for all Libyan teachers and students.
4. Seeking technical support of ICT and E-learning from successful institutions in other countries by making training programs to develop the local staff.
5. Encouraging the competent authorities in Libya to initiate intensive awareness programs to raise awareness about the importance of E-learning.

**ACKNOWLEDGEMENT**

This work was supported mainly by Al Asmarya Islamic University, Libya; University of Tunis El Manar, Tunisia; and Tallinn University of Technology, Estonia. Thanks to Universiti Putra Malaysia for the full technical and financial support.

**REFERENCES**

[1] M. Khan, S. Hossain, M. Hasan, and C. K. Clement, "Barriers to the introduction of ICT into education in developing countries: The example of Bangladesh.," *Online Submiss.*, vol. 5, no. 2, pp. 61–80, 2012.

[2] A. A. Mirza and M. Al-Abdulkareem, "Models of e-learning adopted in the Middle East," *Appl. Comput. informatics*, vol. 9, no. 2, pp. 83–93, 2011.

[3] A. Hamdy, "Survey of ICT and education in Africa: Libya Country Report Libya-ICT in Education in Libya www.infodev.org/en." Document, 2007.

[4] A. Othman, C. Pislaru, T. Kenan, and A. Impes, "Attitudes of Libyan students towards ICT's applications and e-learning in the UK," in *The Fourth International Conference on e-Learning (ICEL2013)*, 2013, pp. 123–129.

[5] J. K. Tarus, D. Gichoya, and A. Muumbo, "Challenges of implementing e-learning in Kenya: A case of Kenyan public universities," *Int. Rev. Res. Open Distrib. Learn.*, vol. 16, no. 1, 2015.

[6] M. Al-Shboul, "The level of e-Learning integration at The University of Jordan: Challenges and opportunities," *Int. Educ. Stud.*, vol. 6, no. 4, p. 93, 2013.

[7] R. E. Derouin, B. A. Fritzsche, and E. Salas, "E-learning in organizations," *J. Manage.*, vol. 31, no. 6, pp. 920–940, 2005.

[8] A. Rhema and I. Miliszewska, "Towards e-learning in higher education in Libya," *Issues Informing Sci. Inf. Technol.*, vol. 7, no. 1, pp. 423–437, 2010.

[9] G.-J. Hwang, "Definition, framework and research issues of smart learning environments-a context-aware ubiquitous learning perspective," *Smart Learn. Environ.*, vol. 1, no. 1, p. 4, 2014.

[10] I.-C. Hung, X.-J. Yang, W.-C. Fang, G.-J. Hwang, and N.-S. Chen, "A context-aware video prompt approach to improving students' in-field reflection levels," *Comput. Educ.*, vol. 70, pp. 80–91, 2014.

[11] P.-H. Wu, G.-J. Hwang, and W.-H. Tsai, "An expert system-based context-aware ubiquitous learning approach for conducting science learning activities.," *J. Educ. Technol. Soc.*, vol. 16, no. 4, 2013.

[12] A. Meiki, M. Nicolas, M. Khairallah, and O. Adra, "Information and Communications Technology Use as a Catalyst for the Professional Development: Perceptions of Tertiary Level Faculty.," *Int. J. Educ. Dev. Using Inf. Commun. Technol.*, vol. 13, no. 3, pp. 128–144, 2017.

[13] A. E. Johnson, "A nursing faculty's transition to teaching online," *Nurs. Educ. Perspect.*, vol. 29, no. 1, pp. 17–22, 2008.

[14] A. Vyas, "A Journey Beyond: Tracking a Mathematics Teacher Educator's Transition to an Online Instructor," in *E-Learn: World Conference on E-Learning in Corporate, Government, Healthcare, and Higher Education*, 2010, pp. 1257–1262.

[15] A. Andersson and Å. Grönlund, "A conceptual framework for e-learning in developing countries: A critical review of research challenges," *Electron. J. Inf. Syst. Dev. Ctries.*, vol. 38, no. 1, pp. 1–16, 2009.

[16] K. Chiasson, K. Terras, and K. Smart, "Faculty's Transformative Process from Traditional to Online Instruction," in *Society for Information Technology & Teacher Education International Conference*, 2013, pp. 309–314.

[17] J. Mapuva, "Confronting challenges to e-learning in higher education institutions," *Int. J. Educ. Dev. Using ICT*, vol. 5, no. 3, pp. 101–114, 2009.

[18] A. Rhema and I. Miliszewska, "Reflections on a trial implementation of an e-learning solution in a Libyan university," *Issues informing Sci. Inf. Technol.*, vol. 8, no. unknown, pp. 61–76, 2011.

[19] A. Elzawi and S. Wade, "Barriers to ICT adoption in quality of engineering research in Libya: how to bridge the digital divide?," University of Huddersfield, 2012.

[20] A. Al-Azawei, P. Parslow, and K. Lundqvist, "Barriers and opportunities of e-learning implementation in Iraq: A case of public universities," *Int. Rev. Res. Open Distrib. Learn.*, vol. 17, no. 5, 2016.

[21] B. E. Zamani, A. Esfijani, and S. M. Abdellahi Damaneh, "Major barriers for participating in online teaching in developing countries from Iranian faculty members' perspectives.," *Australas. J. Educ. Technol.*, vol. 32, no. 3, 2016.

[22] B. H. Khan, "A framework for e-learning," *bookstoread.com*, 2012. [Online]. Available: http://www.bookstoread.com/framework.

[23] A. M. A. Alsanousi and others, "The Effect of Higher Education quality on Economic Growth in Libya," *Int. J. Acad. Res. Bus. Soc. Sci.*, vol. 7, no. 3, pp. 139–149, 2017.

[24] A. Motivans, T. Smith, and M. Bruneforth, "Teachers and educational quality": Monitoring global needs for 2015.," 2006.

[25] A. Macpherson, G. Homan, and K. Wilkinson, "The implementation and use of e-learning in the corporate university," *J. Work. Learn.*, vol. 17, no. 1/2, pp. 33–48, 2005.

[26] A. Sife, E. Lwoga, and C. Sanga, "New technologies for teaching and learning: Challenges for higher learning institutions in developing countries," *Int. J. Educ. Dev. using ICT*, vol. 3, no. 2, pp. 57–67, 2007.